\documentstyle[preprint,aps]{revtex}
\begin{document}
\draft
\title
{\bf \Large{Evolution of the electronic structure of cyclic 
polythiophene upon bipolaron doping}}
\author{D. Giri and K. Kundu}
\address{Institute of Physics, Bhubaneswar 751005, India}
\date{\today}
\maketitle
\begin{abstract}
Electronic structures of undoped and doped cyclic polythiophene 
(PT) are studied using modified $\sigma$-bond compressibility model. 
Cyclic PT doped with odd number of bipolarons creates an aromatic 
polyene backbone containing (4$n$+2) $\pi$-electrons and the system 
is driven towards the quinoid form. Consequently, we find an 
insulator-metal transition for dopant concentration $\geq$ 
14 mol $\%$ and a $\sim$ 0.8 eV redshift in Fermi energy at 30 
mol$\%$. For even number of bipolarons, we propose 
here that the form having two singly occupied degenerate 
orbitals will be stable in a sufficiently large cyclic PT. 
\end{abstract}
\narrowtext
Polythiophene (PT) and its derivatives
\cite{1,2,3,4,5,6,7,8,9,10,11,12} form an important class 
of organic conducting polymers for certain practical advantages. 
For example, these can be synthesized easily and can be 
doped with dopants\cite{2,3,4}, like $\rm Cl O_4^{-}$, $\rm As F_5$, 
$\rm NOPF_6$ etc. These compounds are also chemically stable in air 
and moisture both in doped and undoped states. Ultraviolet 
photo-electron spectra (UPS) \cite{3} of neutral polythiophene 
show existence of $\pi$-bands in the system and the Fermi energy, 
$E_F$ is $\sim 1$ eV above the valence band (VB). So, neutral PT 
is an intrinsic semiconductor with a band gap of $\sim 2$ eV between 
the valence band and conduction band (CB). However, metallicity 
can be induced in PT by doping. For example, high quality films 
of PT obtained by electro-chemical polymerization of dithiophene 
\cite{2} under mild conditions exhibit a metallic transition 
with $\rm Cl O_4^{-}$ doping at $\sim 20$ mol $\%$ doping level. 
The electrical conductivity of doped PT, however, shows wide 
variation depending on the method of preparation and dopants. 
UPS of PT films\cite{4} also reveals a finite density of states 
at $E_F$ at saturation doping (30 mol $\%$ doping with $\rm Cl 
O_4^{-}$). Another important finding from UPS is the 
redshift\cite{3,4} of $E_F$ by $\sim 0.8$ eV at this doping. 
However, the mechanism for the insulator to metal transition 
in PT by doping is an open and polemical question. 

Since the ground state of neutral PT is non-degenerate, dopants 
induce polaronic and bipolaronic defect states\cite{8} in the 
system. At low doping concentrations, two optical transitions  
at $\sim 0.4$ and $\sim 1.3$ eV are due to polaronic defects 
states\cite{5,8}. At intermediate doping level, these transitions 
are replaced by two other transitions\cite{5,8} at $\sim 0.6$ 
and $\sim 1.5$ eV. These are bipolaronic transitions. 
Another interesting result is that X-ray photoelectron spectra 
(XPS) of C1s of PT \cite{4} at saturation doping show an intense 
peak at $\sim 286$ eV. This peak shows complete reversibility 
during doping and undoping cycle. So, this is an enhanced 
$\pi - \pi^{*}$ shakeup satellite\cite{4} arising from the 
modification of the electronic structure of PT films by
doping. Furthermore, its intensity is not related to the dopant 
concentration (C) indicating the absence of localized positive 
charges in doped PT. This then implies that metallic conductivity in 
PT arises due to uniform extraction of $\pi$-electrons from PT by 
dopants. We further note that in high quality PT films \cite{2}, 
no inter-band optical transition is observed for C $\geq 20 \rm \%$

Since one dimensional open chain polyenes can obtain stabilization 
from extensive bond alternation (Peierls' instability\cite{13}), 
formation of localized charges and a band gap between the VB and 
the bipolaron band, if formed  should be the natural outcome. 
So, the absence of localized positive charges and the appearance 
of metallic conductivity in highly doped PT indicate that some other 
structural factor plays a significant role in the suppression of 
Peierls' instability. This consideration leads us to explore the 
evolution of the electronic structure of cyclic polythiophenes upon 
doping. The motivation is that a planar mono-cyclic polyenes with 
(4$n$ + 2) ($n$ = 0,1, ......) $\pi$-electrons have innate aromatic 
stabilization\cite{14}. On the other hand the same system with 4$n$ 
$\pi$-electrons are anti-aromatic. So, the innate aromatic 
stabilization of (4$n$ + 2) $\pi$-electrons ring can play 
fundamentally important role in the insulator to metal 
transition in doped PT. Other important considerations are that 
a ring has the desired translational invariance of the infinite 
system and its energy spectra is  comparable to that of the 
infinite system.  

In our model calculation we assume that PT ring can be treated as 
a planer conjugated polyene. The minimum number of thiophene units  
required to form a planer PT ring is twelve \cite{12}. 
Electronic structures of undoped as well as doped PT rings are 
calculated within the framework of modified $\sigma$-bond 
compressibility model \cite{14} which has been discussed 
thoroughly in Ref.9. The total Hamiltonian ($H$) in this formalism is 
\begin{eqnarray}
H & = & \sum_{p} \epsilon_{p} a_{p}^{\dag} a_{p} + \frac{1}{2} 
\sum_{<p q>} (V_{pq} a_{p}^{\dag} a_{q} + h.c.) \nonumber \\
& + & \frac{C'}{2} \sum_{<p q>} 
\sum V_{p q} (R_{p q} - R_{p q}^{0} + B). 
\end{eqnarray}
Here, $<p q>$ denotes summation over those atoms which are joined 
by a $\sigma$ bond  (nearest neighbor). 
In (1), $a_{p} (a_{p}^{\dag})$ destroys (creates) an electron 
at the $p$-th site. $\epsilon_p$, the self energy of the electron 
at the $p$-th site and the modified $\sigma$-bond compressibility 
model assume that 
\begin{eqnarray}
\epsilon_{p} & = & \epsilon_{p}^{0} - A \chi_{p}\sum_{q} [ (R_{p q} 
- R_{p q}^{0} + B) e^{-R_{p q}/B} \nonumber \\
& - & (R_{p q}^{ref} - R_{p q}^{0} + B) e^{-R_{p q}^{ref}/B} ]
\end{eqnarray}
Summation in (2) spans over all atoms bonded to the $p$-th 
atom. $R_{p q}^{0}$ is the theoretical $\sigma$ bond 
length\cite{14} of the $(p, q)$ bond. $R_{p q}^{ref}$ denotes some 
suitably chosen reference bond length of the $(p, q)$ bond and the 
standard literature values\cite{6,7,9,14} of $\sigma$ bond 
lengths in neutral PT are used for these parameters. ${\chi_{p}}$ 
is the coupling parameter for the $p$-th site arising due to the 
coupling of the electronic structure to the lattice deformation. 
$V_{p,q}$ is the tunneling matrix element that allows the electron 
at the $p$-th site to tunnel to the $\sigma$-bonded $q$-th site. 
We take $V_{p,q} = - A e^{-R_{p,q}/B}$, where $A$ and $B$ are the 
scaling parameters for the energy and the length respectively 
(ref. 14). The last term in $H$ defines the elastic energy 
of the $(p, q)$  $\sigma$ bond. Here $C'$ is another adjustable 
parameter which we assume to be independent of the nature of the bonds. 

The parameters $\{\chi_{p}\}$, $\epsilon_{p}^{0}$, $A$, $B$, $C'$ 
are so chosen as to obtain the experimentally measured band 
gap ($E_g$) between the VB and the CB in the neutral PT, the 
theoretically calculated valence band width and the ground state 
geometry of neutral PT which is comparable to 
the geometry obtained by other standard methods \cite{6,7,8,9,10,11}. 
Values of these parameters are found to be identical to those 
obtained previously for PT chains in Ref.9, yielding a band gap 
of $\sim$ 2 eV and a valence band width of $\sim$ 2.77 eV. 
In addition, a single bipolaron defect  creates
two states at $\sim$ 0.56 and $\sim$ 1.436 eV above the VB.  
So, bipolaronic transitions also compare well with experimental 
values\cite{5,7,8}. Furthermore, total +2 charge of the bipolaronic 
defect spreads over approximately ten thiophene units and 75$\%$ 
of the total charge resides over five units. So, for a single 
bipolaronic defect states, results do not depend significantly on 
the boundary condition.

Energy levels in neutral PT separate into five well defined bands. 
The lower-most and the uppermost levels in each band are non-degenerate 
while other levels in any given band are doubly degenerate. So, 
odd number (2$l$-1, $l$=1,2.......) of bipolarons can be introduced 
in PT ring by removing one pair of electrons from the uppermost level 
and the rest from the subsequent ($l$-1) pairs of states and the 
configuration is unique. Since each thiophene unit has four 
$\pi$-electrons in its carbon backbone, the number of 
$\pi$-electrons in the polyene backbone of a PT ring containing 
N thiophene rings is 4N. But due to incorporation of (2$l$-1) 
bipolarons, the polyene backbone in the doped system converts to 
a [4(N-$l$) + 2] or (4$n$+2) $\pi$-electron system. On the other 
hand, for even number (2$l$) of bipolarons, the number of 
$\pi$-electrons in the polyene backbone is 4(N-$l$) or 4$n$. 
Furthermore, when a PT ring contains even number of bipolarons, 
the occupation of the last pair of degenerate levels can 
be either (2,0)/(0,2) or (1,1). In subsequent discussion 
these configurations will be referred to as S and T respectively.

The dependence of the gap between the highest occupied and the 
lowest unoccupied orbitals (HOMO and LUMO respectively) on 
C in a cyclic PT containing hundred thiophene 
units is shown in Fig.1. The evolution of the HOMO-LUMO gap in the 
open PT chain of same length is also included for comparison. We 
first discuss the case of odd bipolarons. While the periodicity in 
the open chain doped PT depends strongly on C, for this case, the 
the unit cell  of the doped system is found to be  the thiophene ring 
as C increases beyond a critical value (C $\geq$ 14 mol$\%$).  
The structure of the modified thiophene ring for some representative 
dopant concentrations above the critical value and also the 
structure of this ring in neutral PT are given in Fig.2. We note that
the bond alternation in the polyene backbone of PT reduces 
significantly due to doping above the critical dopant concentration. 
So, we assume that if the bipolaronic doping creates a (4$n$+2) 
$\pi$-electrons polyene backbone in PT, bands in the mother system 
undergo reorganization and a new CB ensues in the process. We 
further assume that energy levels in the new CB can be 
obtained from a Tight binding model with nearest neighbor hopping 
only. So, we write $E(k) = (B/2) \cos k$ where $B$ is the band 
width of the CB and as usual $k = \frac{2 \pi m}{\rm N}$. Then 
for large N we obtain 
\begin{equation}
\Delta E = E_{\rm LUMO} ({\rm N}) - E_{\rm HOMO} ({\rm N}) \simeq 
\frac{B \pi}{{\rm N}} \sin k
\end{equation}
where $k = (k_{\rm HOMO} + k_{\rm LUMO})/2$. 
We now consider two PT rings with $\rm N_1$ and $\rm N_2$ units 
respectively. These rings are doped with odd number of bipolarons 
but with same C. From Eq.(3) we then obtain 
\begin{equation}
{\rm N_2} = \frac{\Delta E ({\rm N_1})}{\Delta E ({\rm N_2})} \rm N_1
\end{equation}
Our results are given in Table I and the agreement is definitely 
excellent. We also show in Table I the Fermi energy, 
$E_F$ [=$(E_{\rm HOMO} + E_{\rm LUMO})/2$] for two different 
values of N for a given C and also for different C. It should
be noted that $E_F$, particularly at higher dopant concentrations, 
does not show any appreciable dependence on N. This further shows 
the correctness of our hypothesis.  The change of $E_F$, 
$\Delta E_F = \mid E_F({\rm doped}) - E_F({\rm neutral})\mid$ 
is shown in Fig.3. At $C = 30$ mol$\%$ we obtain 
$\Delta E_F \sim 0.84$ eV, while the experimental value\cite{4} 
is $\sim 0.8$ eV. Hence, our theory also successfully predicts 
$\Delta E_F$.

We consider now the case of doping the system with even number of 
bipolarons. Since two uppermost degenerate levels are singly 
occupied in the T-configuration, the doped system cannot gain 
stabilization by pseudo Jahn-Teller effect \cite{14}. So, the 
evolution of the structure of the T-configuration upon doping 
is similar to the case of doping PT with odd number of bipolarons. 
But due to the interaction of two polarons, the evolution of the 
HOMO-LUMO gap does not strictly follow Eq.(4). This can be seen by 
considering the case of 20 mol $\%$ dopant concentration. 
For ($\rm N_1$, $\rm N_2$, $\rm N_3$) = (80, 100, 120), 
we have ($\Delta E (\rm N_1)$, $\Delta E (\rm N_2)$, 
$\Delta E (\rm N_3)$) $\simeq$ (0.0974, 0.0791, 0.0666). 
From the first two entries we obtain $\rm N_2 \simeq 98.49$ while 
the last two entries yield $\rm N_3 \simeq 118.76$. So, the 
deviation compared to the previous case is quite significant. 
The dependence of $\Delta E_F$ on C for a 100 unit PT ring is also 
shown in Fig.3. We note that the magnitude of $\Delta E_F$ for 
C = 32 mol $\%$ ($\sim$ 0.83 eV) agrees very well with the 
experimental value\cite{4}.

The structure of the S-configuration at a given C is similar to 
the corresponding structure of the doped PT chain. This is the 
manifestation of pseudo Jahn-Teller effect. We further note 
that at C = 20 mol $\%$ and for N = 100 and 140 we obtain a gap 
of $\sim$ 0.55 and $\sim$ 0.54 eV respectively. Similarly, 
for C = 40 mol $\%$ we obtain $\sim$ 0.3 eV for both cases. 
This shows that the HOMO-LUMO gap has reached the saturation value. 
So, the gain in electronic energy due to structural distortion has 
reached the asymptotic limit.  The stabilization energy 
per unit cell ($\Delta E$) of the S-configuration over the 
other configuration is defined as 
$\Delta E_{\rm ST} = \frac{(E_{\rm S} - E_{\rm T})}{\rm N}$. 
Its dependence on C and for two different values of N (100 and 140) 
are shown in Fig.4. It is important to note that $\Delta E_{\rm ST}$ 
reduces with increasing N. For example, at C = 20 mol $\%$ and 
40 mol $\%$, the difference in $\Delta E_{\rm ST}$ is $\sim$ 
-0.13 and -0.18 respectively in the scale of $10^{-3}$ eV. 
Since we are considering values of N which are quite close, 
predictably the difference is also small.

To understand the significance of our results we note 
that in a doped PT chain the unit cell contains more than one 
thiophene ring. So, the formation of a new CB upon doping 
is not possible and a transformation from the bipolaron lattice 
to the polaron lattice\cite{8} has been proposed to explain the 
insulator-metal transition. On the other hand, the aromaticity 
of the polyene backbone of a cyclic PT doped with odd number of 
bipolarons drives the system towards the quinoid form. Since, 
the thiophene ring is again the unit cell in this structure, 
a new CB ensues in the doped system with $E_F$ inside it. 
We further note that the theoretically estimated upper 
limit\cite{14} of $n$ beyond which a neutral cyclic polyene with 
(4$n$+2) $\pi$-electrons would exhibit bond alternation is $\sim$ 8. 
The experimental value\cite{14}, however, is $\sim$ 4. On the other hand, 
values of $n$ are at least an order of magnitude large in our 
calculations. So, we contend that our results for the case of odd 
number of bipolarons will not show any significant deviation due 
to further increase in N. 

For even number of bipolarons we find that the S-configuration is 
stable compared to the other configuration for values of N 
considered here. However, two important results, namely (i) the 
HOMO-LUMO gap has reached the asymptotic value, 
(ii) $\Delta E_{\rm ST}$ decreases with increasing N, point to a 
$\rm N_c$, a critical value of N after which the T-configuration 
will be stable. Furthermore, for $\rm N \rightarrow \infty$, results 
cannot critically depend on the integer character of the number 
of bipolarons. So, we propose that the T-configuration will 
be the stable species for $\rm N > N_c$. Two important experimental 
results, namely the magnitude of $\Delta E_F$ and the absence of 
inter-band transitions in heavily doped PT also indicate such  
possibility. We observe in this connection that the variation in the 
conductance of doped PT may be due to the presence of the 
T-configuration in the sample. So, we conclude that the evolution 
of the electronic structure of a cyclic PT upon doping can explain 
successfully many important experimental results including the 
metal-insulator transition.

\begin{figure}
\caption{The HOMO-LUMO gap (in eV) as a function of mol $\%$ dopant 
concentration: (a) for odd number of bipolarons in a PT ring 
($\diamond$); (b) for T-configuration ($\Box$);  
(c) for S-configuration (+), (d) for open PT chain ($\times$).} 
\end{figure}
\begin{figure}
\caption{The bond lengths (in $\rm \AA$) of the thiophene ring at 
different dopant concentration: (a) at 18 mol $\%$; (b) 26 mol $\%$; 
(c) 30 mol $\%$; (d) 50 mol $\%$. Values within the parenthesis are for 
neutral PT ring.}
\end{figure}
\begin{figure}
\caption{$\Delta E_F$ (in eV) as a function of dopant concentration
(mol $\%$):(a) for odd number of bipolarons ($\Box$); (b) for 
T-configuration (+).}
\end{figure}
\begin{figure}
\caption{$\Delta E_{\rm ST}$ (in eV) as a function of dopant
concentration (mol $\%$):(a) for 100 unit PT ring ($\Box$); 
(b) for 140 unit PT ring (+).}
\end{figure}

\begin{table}
\caption{Calculated values of $\rm N_2$ at different dopant
concentrations (mol $\%$). Values of $E_F$ and $B$ (in eV) at 
different N and also at different C.}
\vspace {0.3in}
\begin{tabular}{|ccccccccccccccccccc|}
\hline
\multicolumn{1}{|c}{$\%$ Defect}&
\multicolumn{1}{c}{}&
\multicolumn{1}{c}{$\rm N_1$}&
\multicolumn{1}{c}{}&
\multicolumn{1}{c}{$\Delta E (\rm N_1)$}&
\multicolumn{1}{c}{}&
\multicolumn{1}{c}{$\rm N_2$}&
\multicolumn{1}{c}{}&
\multicolumn{1}{c}{ $\Delta E (\rm N_2)$}&
\multicolumn{1}{c}{}&
\multicolumn{1}{c}{$\rm N_2$ (calculated)}&
\multicolumn{1}{c}{}&
\multicolumn{1}{c}{$E_F(\rm N_1)$}&
\multicolumn{1}{c}{}&
\multicolumn{1}{c}{$E_F(\rm N_2)$}&
\multicolumn{1}{c}{}&
\multicolumn{1}{c}{$B(\rm N_1)$}&
\multicolumn{1}{c}{}&
\multicolumn{1}{c|}{$B(\rm N_2)$}\\
\hline
15 & & 91 & & 0.06838 & & 117 & & 0.05322 & & 116.924 & & -0.6554 & &
-0.6549 & & 3.2336 & & 3.2334\\
20 & & 90 & & 0.09245 & & 110 & & 0.07566 & & 109.967 & & -0.6344 & &
-0.6342 & & 3.3967 & & 3.3965\\ 
25 & & 88 & & 0.11161 & & 120 & & 0.08186 & & 119.979 & & -0.6743 & &
-0.6741 & & 3.5459 & & 3.5456\\
30 & & 100 & & 0.10522 & & 140 & & 0.07516 & & 139.992 & & -0.7638 &
& -0.7637 & & 3.6492 & & 3.6491\\
40 & & 95 & & -0.11549 & & 125 & & 0.08777 & & 124.998 & & -1.0015 &
& -1.0015 & & 3.7682 & & 3.7682\\
50 & & 100 & & 0.11084 & & 140 & & 0.07917 & & 139.999 & & -1.2629 &
& -1.2629 & & 3.8360 & & 3.8360\\
\hline
\end{tabular}
\end{table}
\end{document}